\documentclass[prd,nofootinbib,a4paper,twocolumn,showpacs,showkeys]{revtex4}

\usepackage{braket}
\usepackage{vector}
\usepackage{mathrsfs}
\usepackage{color}
\usepackage{amssymb, amsmath}
\usepackage{amsfonts}
\usepackage{graphicx,epsfig}
\usepackage{latexsym}
\usepackage{hyperref}

\makeatother

\newcommand{\intg}{{\textstyle\int}\,}

\renewcommand{\Im}{{\rm Im}}
\newcommand{\CP}{\textit{CP}}
\newcommand{\dg}{\mathscr{D}^4}
\renewcommand{\vec}[1]{{\bf #1}}

\newcommand{\hubble}{H}
\newcommand{\invt}{z}

\begin{document}

\title{Quantum corrections to leptogenesis from the gradient expansion}

\author{M. Garny$^{a}$}
\email[\,]{mathias.garny@ph.tum.de}

\author{A. Hohenegger$^{b}$}
\email[\,]{andreas.hohenegger@mpi-hd.mpg.de}

\author{A. Kartavtsev$^{b}$}
\email[\,]{alexander.kartavtsev@mpi-hd.mpg.de}  
 
\affiliation{%
$^a$Technische Universit\"at M\"unchen, James-Franck-Stra\ss e, 85748 Garching,
Germany\\
$^b$Max-Planck-Institut f\"ur Kernphysik, Saupfercheckweg 1, 69117 Heidelberg,
Germany}

\begin{abstract}
  Using the closed-time-path formalism we quantify gradient corrections
  to the kinetic equations for leptogenesis, that are neglected in the 
  standard Boltzmann approach.
  In particular we show that an additional \CP-violating source term arises,
  which is non-zero even when all species are in local
  thermal equilibrium.
  In the early universe it is proportional to the expansion rate and
  would vanish for static equilibrium configurations,
  in accordance with the Sakharov conditions.
  We find that for thermal leptogenesis in a standard
  cosmological background the additional source term is small.
  However, it can become the dominant source in the limit of ultra-strong
  washout.
\end{abstract}

\maketitle

\section{Introduction}\label{sec:Introduction}

Today the observed universe almost entirely consists of matter, i.e.~is 
baryonically asymmetric. An attractive explanation of the 
observed asymmetry is provided by the baryogenesis via leptogenesis 
scenario \cite{Fukugita:1986hr}. In this scenario the Standard Model 
is supplemented by heavy Majorana neutrinos. The \CP- 
and lepton-number violating decay of the heavy neutrinos
produces a net 
lepton asymmetry. The rapid expansion of the universe ensures that this 
asymmetry is not washed out by the inverse decay and scattering processes
\cite{Sakharov:1967dj}. Finally, anomalous electroweak processes 
convert the generated lepton asymmetry to the observed baryon asymmetry 
\cite{'tHooft:1976up,Kuzmin:1985mm}.

The computation of the asymmetry in terms of the neutrino masses and mixing
parameters 
requires a microscopic description of the out-of-equilibrium decay process. The
canonical 
approach is based on Boltzmann equations, furnished with decay and scattering
rates computed 
from the vacuum S-matrix elements. 
The size of various corrections to this approximation 
has been investigated by including e.g.~deviations from kinetic equilibrium, 
quantum statistical terms, thermal masses and matrix 
elements~\cite{Covi:1997dr,Giudice:2003jh,Basbol:2007,HahnWoernle:2009qn,
Garayoa:2009my,Kiessig:2010pr}.

Beyond this it is important to check the validity of the semi-classical
treatment
embodied by the Boltzmann approach. This is particularly relevant for
leptogenesis where loop effects and unstable particles are essential for the
generation of the asymmetry.
Within the standard Boltzmann (bottom-up) approach, these typically give rise to
double counting problems
as well as ambiguities related to the application of equilibrium quantum
field theory for the analysis of out-of-equilibrium processes.

These ambiguities can be resolved within the \textit{nonequilibrium}
closed-time-path or
Schwinger-Keldysh formalism. In contrast to the 
bottom-up approach this may be viewed as a top-down approach.  Here, 
the underlying microscopic description
is based on the full quantum-mechanical evolution equation for the
expectation value of the lepton current. Starting from the latter,
quantum-corrected Boltzmann-like kinetic equations can be derived,
which are inherently free of the double-counting
problem, and include medium corrections to the CP-violating parameters
in a consistent way~\cite{Garny:2009rv,Garny:2009qn,Garny:2010nj},
see also \cite{Buchmuller:2000nd,Beneke:2010wd}. Furthermore, a
possible influence of off-shell and memory effects has also been
studied in this approach~\cite{DeSimone:2007rw,DeSimone:2007pa,Anisimov:2008dz,Anisimov:2010aq}.

The reduction of the full quantum equations of motion to Boltzmann-like
equation requires a so-called gradient expansion in
powers of space-time gradients $\partial_{X^\mu}$~\cite{KB:1962}.
Physically, it amounts to an expansion in powers of the ratio of the microscopic
time-scale $t_{mic}\sim 1/M_i$ and the  macroscopic scales
$t_{mac}\sim 1/\Gamma_i,1/{\hubble}$. This means, it requires that the
decay rates $\Gamma_i$ and the cosmic expansion rate ${\hubble}$ are
much smaller than the corresponding
right-handed neutrino masses $M_i$.

The standard Boltzmann treatment of leptogenesis relies
on the zeroth order in the gradient expansion, i.e.~neglects
all  gradient terms. 
Although the condition $t_{mic}\ll t_{mac}$ 
is typically well fulfilled in thermal
leptogenesis, higher gradient terms could still be important for the calculation
of the \emph{asymmetry}, since the latter depends on the tiny difference between
particle and anti-particle interactions
and densities. At zeroth order t
he deviation from equilibrium, which is
crucial according to the third Sakharov condition, is 
described by the non-equilibrium distribution function of
the heavy Majorana neutrinos: 
$\Delta f_{\psi_i}(p) = f_{\psi_i}(p) - f_{\psi_i}^{eq}(p)$.
Consequently, the \CP-violating source term is proportional to
$\Delta f_{\psi_i}$.

The gradient terms capture another potential source for a
deviation from thermal equilibrium, which is due to the
time-dependence of the effective temperature $T(t)$ of the thermal bath of
leptons, quarks,
gauge- and Higgs bosons.
Consequently, we expect that the gradient \CP-violating source 
term is proportional to $\dot T\equiv dT/dt$. In contrast to the standard
source,
it remains non-zero in the limit $\Delta f_{\psi_i}\rightarrow 0$,
and therefore could be important even though it is suppressed by the
Hubble scale, $\dot T \simeq -HT$. Furthermore, it is conceivable that leptogenesis or baryogenesis
occurs in the early Universe simultaneously with 
other non-equilibrium phenomena such as (p)reheating or phase transitions in
which case gradient contributions could be strongly enhanced. Note that such
gradient terms are crucial for electroweak baryogenesis, see
e.g.~\cite{Kainulainen:2002th,Prokopec:2003pj,Prokopec:2004ic,Konstandin:2004gy,Garbrecht:2003mn,Konstandin:2005cd,Cirigliano:2009yt}.
Finally, on the formal level, it is well-known that including first-order
gradient
terms ensures the validity of exact conservation laws of the kinetic
equations~\cite{Knoll:2001jx}.

The aim of this work is to quantify the leading CP-violating
source term proportional to $\dot T$ arising from
gradient corrections to the kinetic equations.
In section~\ref{sec:Setup}, we briefly review the  Boltzmann and the
closed-time-path approaches and set up our notation. In the same section
we discuss the gradient expansion and the Boltzmann limit.
Next we analyze the new source term and discuss the results.
\begin{itemize}
 \item [(i)] As we demonstrate in section~\ref{sec:Delta}, the additional source
term has a qualitatively new structure and does not vanish even if all species
are in \textit{local} thermal equilibrium. Furthermore, we demonstrate that it
becomes important for very heavy Majorana neutrinos and ultra-strong washout.
\item[(ii)] In section~\ref{sec:Epsilon} we argue that  the gradient corrections
 also modify the well-known standard source term.
\end{itemize}
Finally we summarize our results and conclude in section~\ref{sec:Conclusion}.

\section{Generation of an asymmetry in the CTP approach}\label{sec:Setup}

In this section, we first review the standard Boltzmann approach, and
then the closed-time-path (CTP) approach for the generation of a
$B-L$ asymmetry. In order to illustrate the effect of the gradient
corrections, we consider a toy model which has also been used
in~\cite{Hohenegger:2008zk,Garny:2009rv,Garny:2009qn}.
However, the derivation and the structure of the results are generic,
and similar gradient corrections will also be present in  
phenomenological scenarios such as thermal leptogenesis. The Lagrangian of the
model reads
\begin{align}
\label{lagrangian}
{\cal L}&=\frac12 \partial^\mu\psi_i\partial_\mu\psi_i
-\frac12 M^2_i \psi_i\psi_i
+\partial^\mu \bar{b}\partial_\mu b\nonumber-
m^2\bar{b}b\nonumber\\
&-\frac{\lambda}{2!2!}(\bar{b}b)^2-
\frac{g_i}{2!}\psi_i bb-\frac{g^*_i}{2!}\psi_i \bar{b}\bar{b}
\,,\quad i=1,2\,.
\end{align}
It can be considered as  toy model for a generic baryogenesis
scenario in which the asymmetry is produced by the out-of-equilibrium
decay of some heavy species and where the \CP-asymmetry in the 
decay is induced by the one-loop contributions depicted in Fig.\,\ref{interference}. For example, in thermal
leptogenesis the real scalar fields $\psi_i$ model the heavy Majorana 
(s)neutrinos, whereas the complex field $b$ represents the (s)leptons. In 
GUT baryogenesis  the real scalar fields model the heavy 
bosons and the complex field $b$  the baryons.
In the following, we shall simply refer to $b$ as  toy-baryons and to
$\psi_i$ as  toy-neutrinos.

\subsection{Boltzmann approach}

The standard 
approach is based on generalized Boltzmann equations for the
distribution functions $f_a(X,p)$ of on-shell particle species
$a$~\cite{Kolb:1979qa,Kolb:1990vq},
\begin{align}\label{StandardBE}
p^\alpha {\cal D}_\alpha f_a (p) = {\cal C}_a = {\cal C}_a^{gain}[1+f_a (p)] - {\cal
C}_a^{loss} f_a (p)\,,
\end{align}
where ${\cal D}_\alpha$ is the covariant derivative, ${\cal C}_a$ are
collision integrals comprising gain and loss terms, and we suppress the
space-time coordinate $X$ for brevity.
The latter take decays, inverse decays and scatterings into account, with rates
inferred from the $S$-matrix (\emph{in-out} formalism).
\begin{figure}[t!]
  \begin{center}
    \includegraphics[width=0.9\columnwidth]{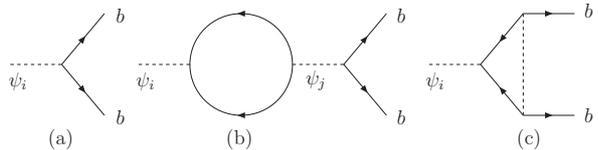}
  \vspace{-4mm}
  \end{center}
  \caption{\label{interference}Tree-level and one-loop-level 
  diagrams of the decay process $\psi_i\rightarrow bb$. The  
  gradient contribution of the vertex diagram is suppressed by higher
  powers of the couplings and is neglected in the following.}
\end{figure} 
If one considers only the decay and inverse
decay processes, see Fig.\,\ref{interference}, then for the toy-baryons  
\begin{subequations}
\label{GainAndLoss}
\begin{align}
\label{Cgain}
{\cal C}_b^{gain}(p)=&\intg d\Pi^3_{k}d\Pi^3_{q}
(2\pi)^4 \delta(k-p-q) |{\cal M}|^2_{\psi_i\to bb}\nonumber\\
&\times \, \,f_{\psi_i}(k)[1+f_b(q)] \,,\\
\label{Closs}
{\cal C}_b^{loss}(p)=&\intg d\Pi^3_{k}d\Pi^3_{q}
(2\pi)^4 \delta(k-p-q) |{\cal M}|^2_{bb \to \psi_i}\nonumber\\
&\times \, \,[1+f_{\psi_i}(k)]f_b (q)\,,
\end{align}
\end{subequations}
where $d\Pi^3_{p}\equiv d^3p/[(2\pi)^3 2E_p]$.
For the antibaryons $f_b$ is replaced by $f_{\bar b}$.
CPT invariance implies for the in-out matrix elements: $|{\cal M}|^2_{\psi_i\to
bb}=|{\cal M}|^2_{\bar b\bar b\to \psi_i}={\textstyle\frac12}|g_i|^2(1+\epsilon^{vac}_i)$ and
$|{\cal M}|^2_{\psi_i\to \bar b\bar b}=|{\cal M}|^2_{bb\to
\psi_i}={\textstyle\frac12}|g_i|^2(1-\epsilon^{vac}_i)$,
where $\epsilon^{vac}_i$ is the usual \CP-violating parameter. As is
well-known, certain
scattering contributions $bb \leftrightarrow \bar b\bar b$ also have to be taken
into
account for consistency within the standard Boltzmann approach, see below.

The total baryon density is given by the time component of the baryon current:
\begin{align}
\label{baryonasymmetry}
  n_B(t) & \equiv V^{-1}{\textstyle \int} dV\, j_0(t,\vec{x})\nonumber\\
& = n_b-n_{\bar b}={\textstyle  \int \frac{d^3p}{(2\pi)^3}}[ f_b-f_{\bar b}]\,.
\end{align}
To stress the analogy with 
phenomenological models, we consider the difference $n_B\rightarrow n_{B-L}
\equiv n_B - n_L$
where $n_L$ (or $n_B$, depending on the interpretation of $b$ as baryons or
leptons, 
respectively) vanishes in the toy model.
Within the standard Boltzmann approach, an evolution equation for $n_{B-L}$
can be obtained by subtracting the Boltzmann equations\,\eqref{StandardBE} for
$f_b$ and $f_{\bar b}$. Dividing the left- and right-hand sides  of
Eq.\,\eqref{StandardBE}  by $p^0=E_p$
and integrating over the momentum space one finds \cite{Kartavtsev:2008fp}:
\begin{align}
\label{EqForBarAsymm}
\frac1{a^3} \frac{d}{dt} \left( a^3 n_{B-L}\right)  ={\textstyle \int}
d\Pi^3_p\,[\,&
{\cal C}_b^{gain}(1+f_b) - {\cal C}_b^{loss} f_b\nonumber\\
-&{\cal C}_{\bar b}^{gain}(1+f_{\bar b}) +{\cal C}_{\bar b}^{loss} f_{\bar
b}\,]\,,
\end{align}
where $a$ is the cosmic scale factor and $t$ the proper time.
Let us stress that, when inserting Eq.\,\eqref{GainAndLoss}, the structure of
the
in-out matrix elements would lead to a generation of an asymmetry even in
equilibrium.
This inconsistency originates from a double counting of decay followed by
inverse decay
and scattering with a $\psi_i$ in the intermediate state. If the quantum
statistical terms are neglected it can be removed by explicitly subtracting the
on-shell part of the s-channel scattering,
a procedure known as real intermediate state subtraction \cite{PhysRevD.56.5431,Pilaftsis:2003gt}.
This is an example for double-counting mentioned in the introduction, and can be
completely resolved in the CTP approach
\cite{Buchmuller:2000nd,Garny:2009rv,Garny:2009qn,Beneke:2010wd}.

The processes which contribute to the right-hand side of 
Eq.\,\eqref{EqForBarAsymm} can be classified as source terms ${\cal S}_0$, which
account for the
generation of an asymmetry, or washout terms ${\cal W}_0$, which tend to deplete
the asymmetry. The subscript should remind the reader that, in the standard
approach, only zero-order gradient contributions are included. 
In the hierarchical limit, $M_2 \gg M_1 \equiv M$, the integrated Boltzmann
equation for the $B-L$ number density in the comoving volume
can be cast into the
form~\cite{Buchmuller:2002rq,Buchmuller:2003gz,Buchmuller:2004nz}
\begin{equation}
\label{StandardRateEq}
  \frac{d Y_{B-L}}{d\invt} = {\cal S}_{0}(\invt) - {\cal W}_{0}(\invt)Y_{B-L}
\,,
\end{equation}
where  $Y=n/s$ is the yield and
$\invt\equiv M/T$ is the inverse temperature normalized by the
mass of the lightest toy-neutrino\footnote{We have replaced the  derivative with
respect to the 
proper time $t$ in
Eq.\,\eqref{StandardRateEq} by the derivative with respect to the dimensionless
inverse temperature 
using the relation $\frac{d}{dt}=\frac{{\hubble}_{|T=M}}{\invt}\frac{d}{d\invt}$
valid in the FRW universe.}.
Performing the usual approximations,
one finds from Eq.\,\eqref{GainAndLoss} that the source term  is
given by
\begin{equation}\label{S0}
{\cal S}_{0}(\invt) =    \epsilon^{vac}\,\kappa \,\invt \gamma_D \,
(Y_{\psi_1} -Y_{\psi_1}^{eq})\,,
\end{equation}
where $\kappa \equiv \Gamma/{\hubble}|_{T = M}$
is the so-called washout parameter
 and the thermally averaged dilation factor 
is given by the ratio of two modified Bessel functions, $\gamma_D\equiv
K_1(\invt)/K_2(\invt)$.
Note that the structure is completely analogous to the phenomenological case.

Equations~\eqref{GainAndLoss} also imply that the washout term is given by 
\begin{equation}
  {\cal W}_{0}(\invt) = \kappa \,\invt \, \gamma_{B-L} \,,
\end{equation}
where $\gamma_{B-L}\equiv \invt^2 K_1(\invt)$.
Note that in a  symmetric configuration, i.e.~for
$Y_{B-L}=0$, the contribution of the washout term
in  Eq.\,\eqref{StandardRateEq} vanishes. We will use this property to calculate
the source term in the
closed-time-path formalism later
on. The yield $Y_{\psi_1}(\invt)$ obeys an equation similar to
Eq.\,\eqref{StandardRateEq},
$d Y_{\psi_1}/d\invt=-\kappa \,\invt \gamma_D \, (Y_{\psi_1} -Y_{\psi_1}^{eq})$.

In the strong washout regime, i.e.~for large $\kappa$, there is a well known
asymptotic 
solution of this system. Up to an overall numerical factor ${\cal O}(1)$ it
reads
\begin{align}
\label{AsymmetryZeroOrder}
\eta_0 \equiv Y_{B-L}(t\to \infty)
\propto \frac{\epsilon^{vac}}{\invt_f}\frac1{\kappa} \,,
\end{align}
where $\invt_f$ is the so-called freeze-out inverse temperature which is
determined 
by the solution of $\kappa \invt_f \gamma_{B-L}=1$.

\subsection{Closed-time-path approach}

Quantum corrections to the semi-classical Boltzmann approach
can be studied using
nonequilibrium quantum field theory techniques which rely on the closed time
path formalism.
In the remainder of this section we briefly review the 
closed time path approach,  its relation to the Boltzmann
approach, and discuss the gradient expansion.

In general the $B-L$ asymmetry \eqref{baryonasymmetry} is given by the
zero-component of the expectation value
of the corresponding quantum mechanical current operator:
\begin{equation}
   j_\mu(x) = 2i\left\langle\left[{\cal D}_\mu b(x)\right]\, \bar b(x) - b(x)\, {\cal
D}_\mu \bar b(x) \right\rangle \,.
\end{equation}
Note that $\langle\cdot\rangle\equiv\mbox{Tr}(\rho\cdot)$, where the density
matrix $\rho$
characterizes the system at some initial time $t_{\rm init}$.
The time-evolution of such \emph{in-in} expectation values
can be described within the closed-time-path or 
Schwinger-Keldysh  approach.
A useful quantity are the Wightman propagators,
\begin{subequations}\label{Wightman}
\begin{align}
  D_>(x,y) & \equiv \bar D_<(y,x) \equiv \langle b(x)\bar b(y) \rangle \,,\\
  D_<(x,y) & \equiv  \bar D_>(y,x) \equiv \langle \bar b(y) b(x) \rangle \,.
\end{align}
\end{subequations}
In terms of these, the $B-L$ current can be expressed as
\begin{equation}
   j_\mu(x) = 2i {\cal D}_{x^\mu} \left[ D_>(x,y) - \bar D_<(x,y) \right]|_{y=x} \,.
\end{equation}
The time-evolution
of the Wightmann propagators is described by so-called Kadanoff-Baym equations,
which are self-consistent Schwinger-Dyson equations formulated on the closed time
path. In the limit of  coinciding arguments they read~\cite{Garny:2009rv,Garny:2009qn}
\begin{align}
\label{KBeqsforD}
\left[\Box_x + m^2\right]D_\gtrless(x,y)|_{y=x}=i& 
 {\textstyle\int\limits^{x^0}_{t_{\rm init}}}\dg z\,
[\Sigma_<(x,z)D_>(z,x)\nonumber\\
&-\Sigma_>(x,z) D_<(z,x)]\,.
\end{align}
 The self-energies $\Sigma_<$ and $\Sigma_>$ in Eq.\,\eqref{KBeqsforD}
can be interpreted as generalizations of gain
and loss terms, respectively~\cite{Garny:2009rv,Garny:2009qn}.
\begin{figure}[t!]
  \begin{center}
    \includegraphics[width=0.30\textwidth]{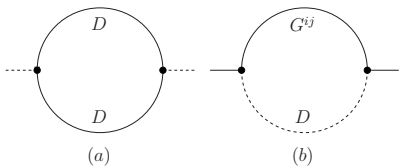}
  \end{center}
  \caption{\label{diagrams} One-loop contributions to the
  self-energies of the toy-neutrinos (a) and toy-baryons (b), respectively.}
\end{figure}
At one-loop level, see Fig.\,\ref{diagrams}, they read
\begin{align}
	\label{Sigmac}
	\Sigma_\gtrless(x,y)&=-g_i g_j^*G^{ij}_\lessgtr(y,x)D_\lessgtr(y,x)\,.
\end{align}
In addition, it is necessary to consider corresponding equations
for anti-particles, differing by $D\to\bar D$ and $\Sigma\to\bar \Sigma$
in Eq.\,\eqref{KBeqsforD}, where the latter are defined analogous to
Eq.\,\eqref{Wightman}.
Let us stress that $D$ and $G^{ij}$ in Eq.\,\eqref{Sigmac} are the full
non-perturbative
Wightmann functions determined by the Schwinger-Dyson equations
for the toy-baryons and toy-neutrinos, respectively.

Using Eq.\,\eqref{KBeqsforD} we can next derive an equation for the divergence
of the baryon current. It reads
  \begin{align}
   \label{DivJ_KB_Derivation}
    {\cal D}^{\mu} & j_\mu(x)  =  2i \left. (\Box_x + m^2) \left[  D_>(x,y) -
\bar D_<(x,y) \right] \right|_{y=x} \nonumber\\
           = & - {\textstyle\int\limits^{x^0}_{t_{\rm init}}} \dg z\, \bigl\{
\Sigma_<(x,z)D_>(z,x)- \Sigma_>(x,z)D_<(z,x)  \nonumber\\
&  - \bar\Sigma_<(x,z)\bar D_>(z,x) + \bar \Sigma_>(x,z)\bar D_<(z,x) \bigr\}
\,.
  \end{align}
Formally, this equation describes the full quantum time-evolution of the
expectation value of the $B-L$ current
starting from arbitrary (Gaussian) initial states at $t=t_{\rm init}$.
Therefore, it is a suitable starting point to derive non-equilibrium
quantum corrections to the standard results.
Note that in the comoving frame in a homogeneous FRW space-time $j_\mu=(n_{B-L},\vec{0})$, i.e.
\begin{equation}\label{K factor}
   {\cal D}^{\mu}  j_\mu(x) = \frac{1}{a^3}\frac{d}{dt}\left(a^3 n_{B-L}\right)
= \frac{1}{K} \frac{d Y_{B-L}}{d\invt}\,,
\end{equation}
where $K^{-1} \equiv s d\invt/dt = sz{\hubble}=s\hubble{|_{T=M}}/z$. Thus,
Eq.\,\eqref{DivJ_KB_Derivation}
indeed constitutes a quantum generalization of Eqs.\,\eqref{EqForBarAsymm} and
\eqref{StandardRateEq}.

Typically, there exists a separation of fast microscopic time-scales,
e.g.~$t_{\rm mic}\sim 1/M$, and the macroscopic evolution characterized by
decay and expansion rates, $t_{\rm mac}\sim 1/\Gamma,1/{\hubble}$.
The crucial observation is that the former sets the scale for
the variation of the two-point functions $D(x,y)$ with respect to the
relative coordinate, given by\footnote{See \cite{PhysRevD.32.1871,Hohenegger:2008zk} for a
proper generalization to curved space-time.} $s = x-y$ ,
while the latter set the scale for the variation with
respect to the central coordinate, given by $X=(x+y)/2$. If there is
a strong hierarchy between microscopic and macroscopic
time-scales, which is typically the case for thermal
leptogenesis, it is possible to expand Eq.\,\eqref{DivJ_KB_Derivation}
in gradients with respect to $X$, and keep only terms up to a
certain order (see e.g. \cite{Muller:PhD}).
For that purpose, it is natural to express the correlation
functions in the Wigner representation, which describes
the ``fast'' variations along the relative coordinate $s$
in momentum space,
\begin{equation}\label{WignerTrafo}
  D_\gtrless(X,p) = {\textstyle \int} d^4 s \, e^{ips}\,
D_\gtrless(X+s/2,X-s/2)\,.
\end{equation}
Then, the gradient expansion of~Eq.\,\eqref{DivJ_KB_Derivation}
in the limit $t_{init}\to-\infty$
can be obtained using the general relation~\cite{Prokopec:2003pj}
\begin{align}
  \label{GradExp}
  {\textstyle \int} \dg z\, A(x,z)& B(z,x)  \nonumber\\
  & = {\textstyle \int {d\Pi_p}} \, e^{-i\diamondsuit} \{A(X,p)\}\{B(X,p)\}\,,
\end{align}
where $d\Pi_p\equiv d^4p/(2\pi)^4$. The derivative operator
\begin{equation}\label{PB}
  \diamondsuit\{.\}\{.\} = \frac12 ( \partial_X^{(1)} \partial_p^{(2)} -
\partial_p^{(1)} \partial_X^{(2)} )\{.\}\{.\}
\end{equation}
generates an expansion in gradients with respect to the ``slow'' coordinate
$\partial_{X^\mu}$
(the superscript refers to the first and second argument).
In particular, Eq.\,\eqref{DivJ_KB_Derivation} can be formally expanded in
powers of $\diamondsuit$:
\begin{eqnarray}\label{DivJ_GradExp}
  {\cal D}^{\mu} j_\mu & = & {\cal D}^{\mu} j_\mu|_0 + {\cal D}^{\mu} j_\mu|_1 +
\dots \,.
\end{eqnarray}

As mentioned above, the standard Boltzmann limit is based on the
zeroth order in the gradient expansion, which reads (see
Appendix~\ref{app:GradExp})  
\begin{align}
 \label{DivJ_Boltzmann}
  {\cal D}^{\mu} j_\mu|_0  =  - {\textstyle \int d\Pi_p} \, \Theta(p_0) \big\{ &
\Sigma_<^0(1+f_b) - \Sigma_>^0\,f_b  \nonumber\\
  - & \bar\Sigma_<^0(1+f_{\bar b}) + \bar\Sigma_>^0\,f_{\bar b} \big\}D_\rho\,.
\end{align}
Note the close similarity of Eqs.\,\eqref{DivJ_Boltzmann}
and~\eqref{EqForBarAsymm}.
The first two terms on the right-hand side can be interpreted as
gain- and loss terms for particles, respectively, while the
last two terms represent gain and loss terms for anti-particles.
Note that anti-baryons have negative baryon-number, hence the relative
minus sign of the second compared to the first line.
Within the CTP approach, the rates of gain and loss processes
are described self-consistently by the self-energies $\Sigma_\gtrless^0$ and
$\bar \Sigma_\gtrless^0$.
Their structure is fixed by the CTP formalism, which resolves the ambiguities of
the
standard Boltzmann approach mentioned before.
Note that in Eq.\,\eqref{DivJ_Boltzmann}, we already introduced the distribution
functions for
baryons, $f_b$, according to the so-called Kadanoff-Baym ansatz,
\begin{align}
\label{Dgtrless}
D_<=f_b\,D_\rho,\quad D_>=(1+f_b)\,D_\rho\,,
\end{align}
together with an analogous relation for anti-baryons involving $f_{\bar b}$.
Here $D_\rho=D_>-D_<$ is the spectral function,
which has a Breit-Wigner-like shape. In the quasi-particle limit it reads
\begin{align}
\label{DrhoQP}
D_{\rho}(X,p)=2\pi\,{\rm sign}(p_0) \,
\delta\left(g^{\mu\nu} p_\mu p_\nu-m^2\right)\,.
\end{align}

The corresponding self-energies in Wigner representation,
see Eq.\,\eqref{WignerTrafo}, read~\cite{Garny:2009rv}
\begin{subequations}
\label{SigmasWT}
\begin{align}
\label{SigmaWT}
\Sigma_\gtrless(X,p)= -\intg d\Pi_{k} & d\Pi_{q} (2\pi)^4
\delta(k-q-p)\nonumber\\
\times & \, g^*_i g_j\,G^{ij}_\gtrless(X,k)D_\lessgtr(X,q)\,,\\
\label{SigmaWTBar}
\bar \Sigma_\gtrless(X,p)= -\intg d\Pi_{k} & d\Pi_{q} (2\pi)^4
\delta(k-q-p)\nonumber\\
\times & \, g_i g^*_j\,G^{ij}_\gtrless(X,k)\bar D_\lessgtr(X,q)\,,
\end{align}
\end{subequations}
where $d\Pi_p\equiv {d^4p}/{(2\pi)^4}$, and 
$G^{ij}_\gtrless$ and $D_\gtrless$ denote the Wightman propagators 
of the toy-neutrinos and the baryons, respectively (see Appendix~\ref{app:G1}
for more details).

The Wightman propagator of the toy-neutrinos is a two-by-two matrix with
non-zero
off-diagonal elements. The latter describe mixing of the heavy fields and, in
particular, 
carry information on the \CP-violation in the system. For a hierarchical mass
spectrum of the 
toy-neutrinos one has \cite{Garny:2009qn}:
\begin{align}
\label{GijFirstOrder}
G^{ij}_\gtrless=\varepsilon_i \, {\cal G}^{ii}_\gtrless+\varepsilon^*_j\, {\cal
G}^{jj}_\gtrless\,, 
\end{align}
so that Eq.\,\eqref{SigmasWT} can be recast in terms of the \CP-violating
parameters,
\begin{align}
\label{CPViolParameters}
\epsilon_i=-2\,\Im\left(g_j/g_i\right) \Im \, \varepsilon_i\,,
\end{align}
and the diagonal components ${\cal G}^{ii}_\gtrless$ of the Wightmann function. 
Using the quasiparticle approximation and integrating Eq.\,\eqref{SigmasWT} over the
time-component
of the four-momenta we obtain gain and loss terms similar to those in
Eq.\,\eqref{GainAndLoss},
see below.

Let us now discuss the \CP-violating source term
within the CTP approach.
As has been mentioned above the contribution of the  washout term vanishes 
in the symmetric configuration, i.e.~for $f_b=f_{\bar b}$, and only the source 
term contributes to the divergence of the baryon current.
Comparing the integrated Boltzmann equation~\eqref{StandardRateEq} for the yield
and Eq.\,\eqref{K factor} suggests to define the \CP-violating source term as
\begin{equation}\label{def of source term}
  {\cal S}(\invt) \equiv K\cdot {\cal D}^\mu j_\mu|_{sym}\,,
\end{equation}
where the subscript means that one evaluates the right-hand side of
Eq.\,\eqref{DivJ_KB_Derivation} for a symmetric configuration.
Using the zero-order gradient contribution \eqref{DivJ_Boltzmann}
and inserting the explicit expressions for the self-energies \eqref{SigmasWT},
we find for the source term in the CTP approach,  to zeroth order in the
gradients: 
\begin{align}
  \label{Zero order source}
  {\cal S}_{0} & \equiv K\cdot{\cal D}^\mu j_\mu|_{0,\,sym} \nonumber\\
  = & 2K\, |g_1|^2  {\textstyle \int} d\Pi_pd\Pi_kd\Pi_q \,
  \Theta(p_0) (2\pi)^4 \delta(k-p-q)\, \nonumber\\
  & \times {\cal G}_\rho^{11}(k) D_\rho(p) D_\rho(q) 
  \cdot \epsilon(k,T) \, \bigl[f_{\psi_1}(k)-
  f_{\psi_1}^{eq}(k)\bigr] \nonumber\\
  & \times
  \bigl([1+f_b^{eq}(p)][1+f_b^{eq}(q)]-f_b^{eq}(p)f_b^{eq}(q)\bigr) \,,
\end{align}
where $f_{\psi_1}$ and ${\cal G}_\rho^{11}$ are the distribution- and spectral
function of the lightest toy-neutrino, respectively,
and we have assumed that the baryons are in equilibrium.
The effective \CP-violating parameter $\epsilon(k,T)=\epsilon^{\rm
vac}+\epsilon^{\rm med}(k,T)$
agrees with the one obtained from the quantum-corrected
Boltzmann equations derived in~\cite{Garny:2009qn,Garny:2009rv} and incorporates
medium corrections~\cite{Garny:2010nj}.
In the limit of an hierarchical mass spectrum \cite{Garny:2009qn,Garny:2009rv}:
\begin{align}
\label{EpsilonHierarchical}
 \epsilon(k,T)  & = \epsilon^{vac} \times \left(1 +  {\textstyle \int
\frac{d\Omega}{4\pi} } [f_b(E_1)+f_{b}(E_2)] \right)\,,
\end{align}
where the second factor accounts for the medium effects and
$E_{1(2)} \equiv {\textstyle\frac12} [(M^2+|\vec{k}|^2)^\frac12 \pm
(1-4m^2/M^2)^\frac12|\vec{k}|\cos\theta]$.

Note also that the structure of the CTP source term automatically
guarantees that no asymmetry is produced in equilibrium.
This means that it is inherently free of the double-counting problem 
and no explicit real intermediate state subtraction is required.
Finally, when neglecting quantum statistical terms and medium corrections,
assuming kinetic equilibrium for $\psi_1$, and inserting the quasi-particle
approximation~\eqref{DrhoQP}, one recovers the standard
source term ${\cal S}_0$ given in Eq.\,\eqref{S0}.

\section{\label{sec:Delta}Additional source term from Hubble expansion}

In this section, we derive and discuss the additional CP-violating source term
for the $B-L$ asymmetry, which is generated due to the time-dependence
of the temperature of the thermal bath, $\dot T=-{\hubble}T$.

An important feature of the CTP approach is that the \CP-violating
source term \eqref{def of source term}
automatically vanishes in complete thermal
equilibrium~\cite{Garny:2009rv,Garny:2009qn}.
Thus, it must be possible to express it in terms of quantities which are
non-zero only out of equilibrium.
Obviously, the deviation of
the toy-neutrino distribution from the equilibrium one,
$\Delta f _{\psi_1}= f_{\psi_1}
- f_{\psi_1}^{eq}$, is such a quantity. As discussed above, a second one 
is given by $\dot T \equiv dT/dt$. Thus, one may decompose the source term
\eqref{def of source term} according to
\begin{eqnarray}
\label{sources}
  {\cal S}(\invt) & \equiv & K\cdot  {\cal D}^\mu j_\mu|_{sym}  \nonumber\\
              & \equiv & {\cal S}(\invt)|_{\Delta f} + {\cal S}(\invt)|_{\dot T}
                    + {\cal S}(\invt)|_{\Delta f \times \dot T} \,. 
\end{eqnarray}
The first contribution is  given by Eq.\,\eqref{Zero order source},
of zeroth order in the gradients,
\begin{equation}
  {\cal S}(\invt)|_{\Delta f} = {\cal S}(\invt)|_0 =K\cdot  {\cal D}^\mu
j_\mu|_{0,\,sym}\,.
\end{equation} 
The second contribution appears only at \emph{first} order in the gradients. It can
be computed by setting all
species into \emph{local} thermal equilibrium (LTE) (i.e.~setting $\Delta f_{\psi_1} =
0$)
with time-dependent temperature $T(t)$,
\begin{equation}
	{\cal S}(\invt)|_{\dot T} = {\cal S}(\invt)|_{1}^{LTE} \equiv K\cdot {\cal
D}^\mu j_\mu|_{1,\,sym}^{LTE}\,.
\end{equation}
Finally, the third contribution requires both $\Delta f_{\psi_1} \not= 0$ and
$\dot T\not=0$.
In the remaining part of this section, we will show that a source term
proportional
to the expansion rate, ${\cal S}\propto {\hubble}$, indeed follows from ${\cal
S}(\invt)|_{\dot T}$.
For a discussion of ${\cal S}|_{\Delta f \times \dot T}$ we refer to
section~\ref{sec:Epsilon}.

The first-order gradient contribution to
the full expression for the $B-L$ current, Eq.\,\eqref{DivJ_KB_Derivation},
can be obtained straightforwardly using Eq.\,\eqref{GradExp}, and keeping
linear terms in the derivative operator (``Poisson bracket'')
defined in Eq.\,\eqref{PB} (see Appendix~\ref{app:GradExp}), 
\begin{align}
\label{DivJ_FirstOrder}
  {\cal D}^{\mu} j_\mu|_1  =   -{\textstyle \int} d\Pi_p & \Theta(p_0)  \bigl( [
\Sigma_<^1(1+f_b) - \Sigma_>^1f_b \nonumber \\
                             - &\bar\Sigma_<^1(1+f_{\bar b}) +
\bar\Sigma_>^1\,f_{\bar b}  ] D_\rho  \nonumber \\
                             +2 &\diamondsuit\{\Sigma_F^0\}\{D_h\} +2
\diamondsuit\{\Sigma_h^0\}\{D_F\} \nonumber\\
                             -2 &\diamondsuit\{\bar\Sigma_F^0\}\{\bar D_h\} -2
\diamondsuit\{\bar \Sigma_h^0\}\{\bar D_F\}  \bigr) \,.
\end{align}
The self-energies $\Sigma_\gtrless^1$ are given by Eq.\,\eqref{SigmasWT} with 
the first-order gradient solutions of the Kadanoff-Baym equations for the
toy-neutrino 
propagator,
\begin{align}
\label{FirstOrderSolution}
\hat G_\gtrless^{1} =  -i
\bigl[ & \diamondsuit \{\hat {\cal G}_R,\hat \varPi^{'}_\gtrless,\hat {\cal
G}_A\}
\nonumber\\
+ &\diamondsuit\{\hat {\cal G}_\gtrless,\hat \varPi^{'}_A,\hat {\cal G}_A\}+
 \diamondsuit\{\hat {\cal G}_R,\hat \varPi^{'}_R,\hat {\cal
G}_\gtrless\}\bigr]\,,
\end{align}
where we have introduced $\diamondsuit\{A,B,C\} \equiv \diamondsuit\{A\}\{BC\} 
+ A\diamondsuit\{B\}\{C\}$ (see Appendix~\ref{app:G1}).
Furthermore, $D_F=(D_> + D_<)/2$, with analogous
relations for self-energies, and $D_h = (D_R + D_A)/2$, involving
retarded and advanced functions (see Appendix~\ref{app:GradExp} for more
details).

As discussed above, the source term ${\cal S}(\invt)|_{\dot T}$ can be obtained
by evaluating Eq.\,\eqref{DivJ_FirstOrder} for a symmetric system in local
thermal
equilibrium. After a somewhat tedious calculation, for which we refer to
Appendix~\ref{app:Delta1}, one finds that at leading order in the toy-neutrino
coupling, and in the hierarchical limit, it is given by
\begin{align}
\label{First order source}
  {\cal S}(\invt)|_{\dot T}
  & \equiv  K\cdot{\cal D}^\mu j_\mu|_{1,\,sym}^{LTE}  \nonumber\\
  &  =      2K\,|g_1|^2 {\textstyle \int} d\Pi_pd\Pi_k d\Pi_q \, \Theta(p_0)
             (2\pi)^4 \delta (k-p-q)\, \nonumber\\
  &        \times{\cal G}_\rho^{11}(k)D_\rho(p) D_\rho(q)  \nonumber\\
  &        \times \epsilon_{\dot T}(k,T) \, \big\{
f_{\psi_1}^{eq}(k)[1+f_b^{eq}(p)][1+f_b^{eq}(q)] \nonumber\\
  &         \qquad\qquad\qquad + [1+f_{\psi_1}^{eq}(k)]f_b^{eq}(p)f_b^{eq}(q)
\big\} \,,
\end{align}
with
\begin{equation}
	\epsilon_{\dot T}(k,T) \equiv  - \epsilon^{vac}  \frac{\dot T}{2T}  
	\left( \frac{\partial}{\partial T} + \frac{u\cdot k}{T} u \cdot \partial_{k}
\right) L_h(k;T),
\end{equation}
and
\begin{align}	
	 L_h(k;T) \equiv & 16\pi {\textstyle \int} d\Pi_\ell \, \Theta(\ell_0) \,
D_\rho(\ell)
	 \left[ {\textstyle \frac12} + f_b^{eq}(\ell) \right]\nonumber\\
      & \times 
	 \left[ \frac{-{\cal P}}{(k-\ell)^2-m^2}
	 +  \frac{-{\cal P}}{(k+\ell)^2-m^2} \right]\,,
\end{align}
where $u=(1,0,0,0)$  in the co-moving frame,
$f_a^{eq}(k)=1/[\exp(u\cdot k/T(t))-1]$, and ${\cal P}$ denotes the principal
value.

This additional source term is the main result of this work.
Note that it is proportional to the expansion rate,
since $\dot T = -{\hubble}T$, as expected from a gradient contribution.
Its structure is \emph{qualitatively} different from the zero-order
source term \eqref{Zero order source}, since it can
be non-zero even when the heavy toy-neutrinos are in equilibrium with the
thermal bath.
We stress that the structure of this source term is, nevertheless, in
agreement with the Sakharov conditions: it is proportional to $\epsilon^{vac}$,
i.e. it requires $CP$ violation, and it requires a deviation from 
equilibrium, which is, however, described by $\dot T$ here.

Let us now discuss the implications of this source term. An explicit
expression for  $\epsilon_{\dot T}$ can be found in Eq.\,\eqref{deltaFull}.
In the strong washout regime most of the asymmetry is generated at 
temperatures $T \lesssim M$. In this limit we obtain for the new \CP-violating
parameter
\begin{equation}\label{deltaNonRel}
	\epsilon_{\dot T}(k,T) \simeq  - \epsilon^{vac} \times\frac{{\hubble}}{2\pi T} \,.
\end{equation}
Inserting $ {\hubble} = 1.66 \sqrt{g_*}\, T^2/M_{Pl} $
with $g_*=106.75$~\cite{Kolb:1990vq}, we find
\begin{equation}
	\epsilon_{\dot T}(k,T) \simeq  - 2.7\, \epsilon^{vac} \times\frac{T}{M_{Pl}} \,.
\end{equation}
This is clearly much smaller than the CP-violating parameter $\epsilon^{vac}$
itself
for typical temperatures $T \sim 10^9{\rm GeV}$.
However, one should keep in mind that the structure of the source term 
\eqref{First order source} differs from the usual one. In particular, if the
neutrino is very close to equilibrium, the zero-order
source term \eqref{Zero order source} is suppressed, which could partly
compensate the smallness of $\epsilon_{\dot T}$. This is typically
relevant in the strong washout regime. 
Therefore, we expect that it is legitimate to neglect the gradient correction 
\eqref{First order source}, unless for extremely strong washout.
In order to estimate its impact, we apply the approximations common
for the standard approach to Eq.\,\eqref{First order source}, and
use Eq.\,\eqref{deltaNonRel}. In this limit Eq.\,\eqref{First order source}
can be re-written as an additional source term to Eq.\,\eqref{StandardRateEq},
\begin{equation}
   \label{AllSourceTerms}
  \frac{d Y_{B-L}}{d\invt} = {\cal S}_0(\invt) + {\cal S}_{\dot T}(\invt) -
{\cal W}_{0}(\invt)Y_{B-L} \,,
\end{equation}
where ${\cal S}_{\dot T}$ is given by
\begin{equation}
{\cal S}_{\dot T}(\invt)
\simeq -\kappa \gamma_{B-L} \frac{{\hubble}|_{T=M}}{M}\frac{\epsilon^{vac}}{2\pi^3} \,.
\end{equation}
Note that we neglected gradient corrections to the washout
term here. This can be justified by the observation that the latter
have the same qualitative structure than the zero-order washout terms,
unlike the source terms.

Let us assume now, that $\kappa$ is very large and the system is very
close 
to thermal equilibrium. In this limit $Y_\psi \simeq Y_\psi^{eq}$ and the
standard 
source term can be neglected. Thus, the rate equation for the baryon number
density 
in the comoving volume takes the form:
\begin{equation}
\label{NewRateEq}
  \frac{d Y_{B-L}}{d\invt} \simeq {\cal S}_{\dot T}(\invt) - {\cal
W}_{0}(\invt)Y_{B-L} \,.
\end{equation}
Using the method of steepest descent we obtain an approximate  solution for 
the asymptotic value of the generated asymmetry:
\begin{align}
\label{AsymmetryFirstOrder}
\eta_{grad} \equiv Y_{B-L}(t\to \infty)
\propto \frac{\epsilon^{vac}}{\invt_f}\frac{M}{M_{Pl}}\,, 
\end{align}
where the freeze-out temperature $\invt_f$ is determined by the same equation
as 
in the zero-order calculation. Comparing this result with
Eq.\,\eqref{AsymmetryZeroOrder}
we see that the asymmetry  is not suppressed by the washout factor $\kappa$ but
on 
the other hand it is strongly suppressed by the ratio of the right-handed
toy-neutrino 
mass to the Planck scale.
The numerical analysis confirms that for large values of the washout parameter 
Eq.\,\eqref{AsymmetryFirstOrder} indeed well approximates the exact result. 

\begin{figure}[t!]
\includegraphics[width=1.0\columnwidth]{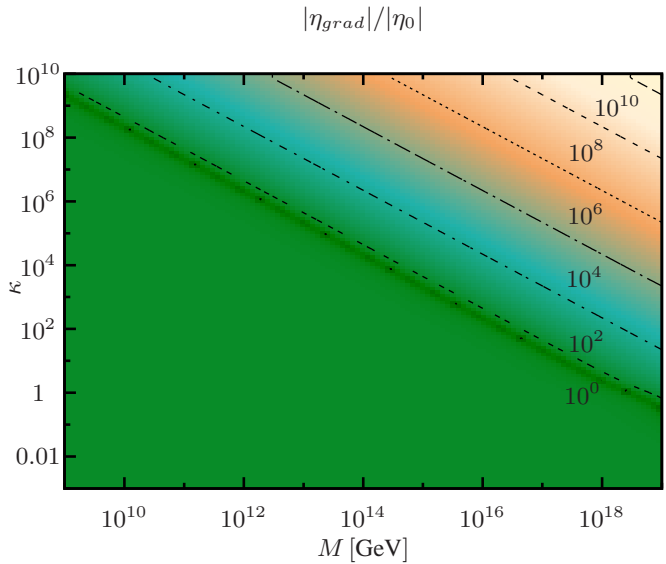}
\caption{\label{Ratio}
Ratio of the final asymmetries $\eta \equiv Y_{B-L}(t \to \infty)$
obtained from the numerical solutions of the rate equations with both 
the gradient and the zero-order source terms taken into account to that with
only the zero-order source term present for various values of the washout
parameter $\kappa$ and the toy-neutrino mass $M$. In the limit $\kappa\to\infty$ the dependence is $|\eta_{grad}|/\eta_0\propto\kappa M/M_{Pl}$.}
\end{figure}

In Fig.\,\ref{Ratio} we present the ratio of numerical solutions of
Eq.\,\eqref{AllSourceTerms} to those of 
Eq.\,\eqref{StandardRateEq}. For a wide range of $\kappa$ and $M$ the gradient
terms are 
subdominant and can safely be neglected. However, in the ultra-strong washout
regime 
and for very heavy Majorana neutrinos they become the dominant source 
of the asymmetry.

\begin{figure}[t!]
\includegraphics[width=1.0\columnwidth]{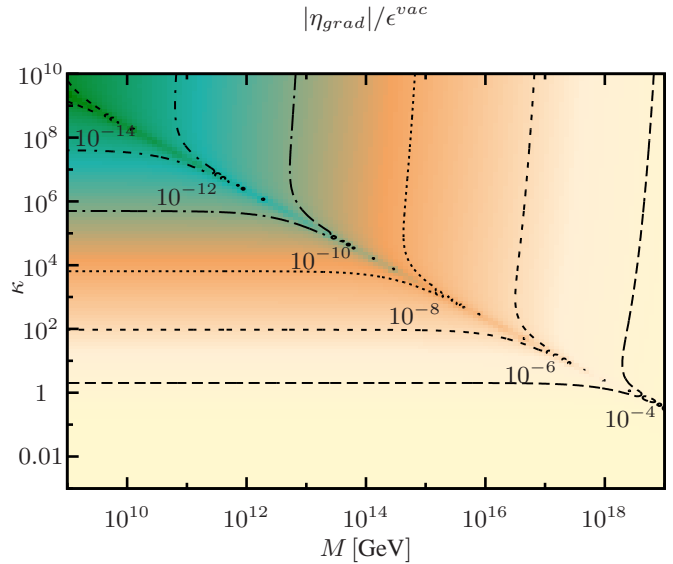}
\caption{\label{NumericalEstimate}Ratio of the generated asymmetry to the
\CP-violating parameter obtained from the numerical solutions of the rate
equations with both the gradient and the zero-order source terms taken
into account. In the upper right part of the plot, the gradient contributions
dominate.}
\end{figure}

Important is not only the relative size of the two contributions but also the
absolute 
value of the generated asymmetry. As is evident from
Fig.\,\ref{NumericalEstimate},
although for large $\kappa$ the contribution of the  gradient source term
dominates,
the efficiency of leptogenesis in the standard cosmological setting
is too small to reproduce the observed value of the asymmetry. However,
the relative importance of the gradient terms could be strongly enhanced if
other non-equilibrium phenomena would occur simultaneously with leptogenesis.
Essentially, the size of gradient terms is determined by the scale of temporal
or spatial inhomogeneities. Since leptogenesis is often regarded to occur
shortly after reheating, it is possible that non-equilibrium fluctuations
occurring for example during (p)reheating are still present. Another possibility
would
be that a phase transition occurs at the temperature relevant for leptogenesis,
possibly related to the breaking of $B-L$ symmetry. If the phase transition
is of first order gradient terms could have a great impact, similarly as
for electroweak baryogenesis. Finally, we note that the analysis presented here
also applies to alternative mechanisms such as 
GUT-scale baryogenesis. Due to the much higher mass
scale of $\sim 10^{16}{\rm GeV}$, the gradient corrections are also larger
in this case.

\section{\label{sec:Epsilon}Gradient correction to \CP-violating parameter}

\begin{figure}[t!]
\begin{center}
\includegraphics[width=1.0\columnwidth]{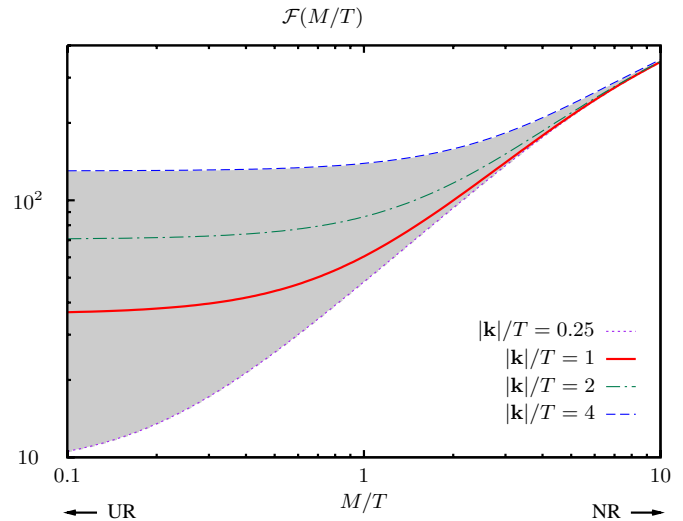}
\end{center}
\caption{\label{F} Dependence of the function ${\cal F}$ on the 
dimensionless inverse temperature for the momenta of the decaying 
particle in the range  $0.25 \le |\vec{k}|/T \le 4$.}
\end{figure}

In this section we discuss the last term in the expansion \eqref{sources}
of the source term. As we will see, it can be
interpreted as a correction to the \CP-violating parameters $\epsilon_i$. As has
been argued
above, $\epsilon_i$ are generated by the off-diagonal components of the
Wightmann propagators
of the toy-neutrinos provided that the latter are complex-valued.
The first-order  solution \eqref{FirstOrderSolution} does have an imaginary
component.
Thus, it also contributes to  the \CP-violation  in the system. Particularly
interesting are the following two terms of Eq.\,\eqref{FirstOrderSolution}:
\begin{align}
\label{GgtrlessCP}
-i
\bigl[\hat {\cal G}_\gtrless \diamondsuit \{\hat \varPi^{'}_A\}\{\hat {\cal
G}_A\}+
\diamondsuit\{ \hat {\cal G}_R\}\{\hat \varPi^{'}_R\} \hat {\cal
G}_\gtrless\bigr]\,,
\end{align}	
because they satisfy the condition \eqref{GijFirstOrder}.
To evaluate them we need explicit expressions for the 
self-energies and the propagators. In a state with zero (or small)
asymmetry the self-energies corresponding to Fig.\,\ref{diagrams}a
read \cite{Garny:2009qn}
\begin{align}
\label{PiRADecomposition}
\hspace{-2mm}
\Pi^{ij}_{R(A)}=\Pi^{ij}_{h}& \pm {\textstyle\frac{i}2}\Pi^{ij}_{\rho}
={\textstyle\frac{-1}{16\pi}}(g_i^*g_j+g_i
g_j^*)[L_h+{\textstyle\frac{i}2}L_\rho],
\end{align}
where the functions $L_h$ and $L_\rho$ are defined in Appendix~\ref{app:G1}.
It has also been shown in \cite{Garny:2009qn} that the diagonal
retarded and advanced propagators may be split into two real-valued diagonal 
matrices:
\begin{align}
\label{GRADecomposition}
{\cal G}^{jj}_{R(A)} =  {\cal G}^{jj}_h \pm
 {\textstyle \frac{i}2}  {\cal G}^{jj}_\rho\,, 
\end{align}
where the off-shell diagonal propagator and the spectral function are given 
respectively by \cite{Hohenegger:2008zk, Garny:2009qn}
\begin{subequations}
\begin{align}
\label{GhReal}
{\cal G}^{jj}_h=-\frac{k^2-M_i^2-\Pi_h^{jj}}{(k^2-M_j^2-\Pi_h^{jj})^2+
{\textstyle\frac14}(\Pi_\rho^{jj})^2}\,,\\
{\cal G}^{jj}_\rho=-
\frac{\Pi_\rho^{jj}}{(k^2-M_j^2-\Pi_h^{jj})^2+
{\textstyle\frac14}(\Pi_\rho^{jj})^2}\,.
\end{align}
\end{subequations}
Substituting Eqs.\,\eqref{PiRADecomposition} and \eqref{GRADecomposition} into 
Eq.\,\eqref{GgtrlessCP} we find 
\begin{subequations}
\begin{align}
\Im\, \varepsilon_i=-\diamondsuit \{\Pi_h^{ij}\}\{{\cal G}^{jj}_h \}
+{\textstyle\frac14}\diamondsuit \{\Pi_\rho^{ij}\}\{{\cal G}^{jj}_\rho \}\,.
\end{align}
\end{subequations}
In the hierarchical case to leading order:
\begin{subequations}
\begin{align}
\label{PihPB}
\diamondsuit \{\Pi_h^{ij}\}\{{\cal G}^{jj}_h \} & \simeq  [\,k^\alpha {\cal
D}_\alpha \Pi_h^{ij}]
 \,{\cal G}^{jj}_h\, {\cal G}^{jj}_h\,,\\
\label{PirhoPB}
\diamondsuit \{\Pi_\rho^{ij}\}\{{\cal G}^{jj}_\rho \} & \simeq 2 [\,k^\alpha
{\cal D}_\alpha 
\Pi_\rho^{ij}]\,
 {\cal G}^{jj}_h\, {\cal G}^{jj}_\rho\,.
\end{align}
\end{subequations}
The decomposition coefficient $\varepsilon_i$ must be evaluated on 
the mass shell of the corresponding quasiparticle. Since we 
assume strong hierarchy of the masses here, ${\cal G}^{jj}_\rho$ 
evaluated on the mass shell of the i'th quasiparticle is negligibly
small for $i\not= j$. Thus, the contribution of Eq.\,\eqref{PirhoPB} can be neglected.
Evaluating the contribution of Eq.\,\eqref{PihPB} we obtain for the 
gradient correction to the
\CP-violating parameter:\footnote{This effect exists also for 
the vertex contribution to the \CP-violating parameter. However, there
it is of a higher order in the coupling constants and for this reason we 
do not consider it here.}
\begin{align}
\label{EpsilonNew}
\epsilon_{\Delta f \times \dot T} &=  -\epsilon_i^{vac}\cdot \frac{2[\,k^\alpha {\cal
D}_\alpha 
L_h]}{M_j^2-M_i^2}\,.
\end{align}
In the strong washout regime the toy-baryons are very close to 
equilibrium so that $L_h(X,p)$ depends on time only through
the dependence of the temperature $T$ on time. Using once again the relation 
${\dot T}=-{\hubble} T$ we find
\begin{align}
\hspace{-1mm}
k^\alpha {\cal D}_\alpha L_h&=\left(E{\textstyle\frac{\partial}{\partial t}}
-{\hubble} \vec{k}^2 {\textstyle\frac{\partial}{\partial
E}}\right)L_h\nonumber\\
&={\hubble}T\bigl[\bigl({\textstyle\frac{E}{T^2}}{\textstyle\frac{\partial}{
\partial (1/T)}}-{\textstyle\frac{\vec{k}^2}{T}} {\textstyle\frac{\partial}
{\partial E}}\bigr)L_h\bigr]\equiv {\hubble}T\cdot F\,,
\end{align}
where $E$ and $\vec k$ are the energy and momentum of the decaying 
toy-neutrino and $F$ is a dimensionless function of T, $\vec k$
and $M_i$. In agreement with our expectations 
the correction to the \CP-violating parameter
is proportional to the Hubble parameter $\hubble$. Evaluating
Eq.\,\eqref{EpsilonNew} in the radiation 
dominated universe we obtain in the hierarchical case :
\begin{align}
\label{Epsilon1Estimate}
\epsilon_{\Delta f \times \dot T}= - \epsilon^{vac}\cdot\frac{M_1}{M_{Pl}}\cdot
\frac{M_1^2}{M_2^2}\cdot \frac{{\cal F}}{\invt^3} \,,
\end{align}
where we have introduced ${\cal F}\equiv 2\cdot 1.66 g_*^\frac12\cdot F$
for convenience. Let us analyze Eq.\,\eqref{Epsilon1Estimate} term by term.
Since most of the asymmetry is generated at $T\sim M_1$, at the 
epoch of leptogenesis $\invt\sim 1$. 
The dependence of the function ${\cal F}$ on the dimensionless 
inverse temperature $\invt$ for various momenta of the lightest toy-neutrino
is presented in Fig.\,\ref{F}.
Roughly speaking ${\cal F}\sim \mathcal{O}(10)$, i.e.~it is of
the order of $2\cdot 1.66 \sqrt{g_*}$.
For hierarchical mass spectrum, $M_1^2/M_2^2 \lesssim 0.1$, the requirement 
of successful  leptogenesis implies that the mass of the lightest 
right-handed neutrino should lie in the range $M_1=10^9-10^{11}$ GeV. 
Even if we take the larger value, $M_1\sim 10^{11}$ GeV, its relative size
compared to 
the Planck mass, $M_{Pl}=1.2 \cdot 10^{19}$ GeV, is still very small, 
of order of $10^{-8}$.
Consequently, the new contribution to the 
\CP-violating parameter is strongly suppressed, primarily due to the 
smallness of the mass of the lightest right-handed neutrino as compared 
to the Planck mass. Let us note again that in GUT baryogenesis scenarios 
where the generation of the asymmetry takes place at higher 
temperatures, $T\sim M_{GUT}$, the relative suppression of the 
new contributions would be less pronounced.
 
\section{\label{sec:Conclusion}Summary and Outlook}

In this paper we have derived gradient corrections 
to the kinetic equations for leptogenesis,
that are neglected in the standard Boltzmann approach.

We have found that there is an additional \CP-violating
source term ${\cal S}_{\dot T}$ with a qualitatively new structure, which
arises due to the time-dependence of the effective temperature of the thermal bath.
It does not vanish even if all particle species  are in \textit{local}
thermal equilibrium. For a standard cosmological background it is
comparable to the conventional one only if the washout parameter
$\kappa\sim M_{Pl}/M_1$ and can therefore be safely neglected in standard
thermal leptogenesis. However, it becomes dominant in the limit of
ultra-strong washout  and for very heavy Majorana
neutrinos, and can play an important role for alternative
baryogenesis mechanisms operating at very high scales, e.g.~at
the GUT scale.

We have also analyzed a contribution to the effective in-medium \CP-violating
parameter which is induced by the gradient terms. Just like the 
${\cal S}_{\dot T}$ term, the new contribution to the \CP-violating parameter
is suppressed by the small ratio of the heavy neutrino mass to the Planck scale.

The gradient terms could be greatly enhanced in a non-thermal
environment, where large temporal or spatial gradients can occur. This may
be relevant when the reheating temperature is very close to the right-handed
neutrino mass, as is often required to avoid the overproduction of gravitinos
in supersymmetric scenarios. Another conceivable situation is that the seesaw
scale is associated to the breaking of a symmetry, possibly $B-L$, in which case
a phase transition could occur at temperatures relevant for leptogenesis.
If it is of first order, gradient terms can play a major role,
similar to electroweak baryogenesis. The additional gradient source term
could then even allow to lower the scale of leptogenesis without
having to rely on resonance effects. This is left for future work.

\subsection*{Acknowledgements}
\noindent
This work was supported by the ``Sonderforschungsbereich'' TR27 and
by the ``cluster of excellence Origin and Structure of the Universe''.
We would like to thank M.\,M.\,M\"uller for sharing his insights on
non-equilibrium field theory. We also
thank S.\,Mendizabal and C.\,Kiessig for useful discussions.

\begin{appendix}

\section{\label{app:GradExp}Gradient Expansion}

In this appendix we derive the gradient expansion of the CTP evolution equation
Eq.\,\eqref{DivJ_KB_Derivation} for the $B-L$ current, following the
lines of \cite{Prokopec:2003pj,Muller:PhD,Muller:2002fh,Berges:2002wt}\footnote{
For simplicity, we will work in flat space-time here. The relevant equations
can be easily generalized to FRW space-time using the results of
ref.~\cite{Hohenegger:2008zk}.}.
For that purpose, it is convenient to switch from the Wightman
functions to an equivalent representation in terms of the
so-called statistical propagator $D_F$ and the spectral function $D_\rho$,
\begin{subequations}
\begin{eqnarray}\label{FRho}
  D_F(x,y)    & \equiv & {\textstyle \frac12} \left[ D_>(x,y) + D_<(x,y) \right]
\,,\\
  D_\rho(x,y) & \equiv & i \left[ D_>(x,y) - D_<(x,y) \right] \,.
\end{eqnarray}
\end{subequations}
Using analogous definitions for $\bar D$ and the self-energies,
Eq.\,\eqref{DivJ_KB_Derivation} can equivalently be written as
\begin{align}\label{DivJ_KB}
    {\cal D}^\mu j_\mu(x) = i {\textstyle \int} & \dg z \, 
\Theta(x^0-z^0)\Theta(z^0-t_{init}) \\
         \times  \! \bigl[ & \Sigma_F(x,z)D_\rho(z,x) - \Sigma_\rho(x,z)D_F(z,x)
\nonumber\\
                 {} - & \bar\Sigma_F(x,z)\bar D_\rho(z,x) + \bar
\Sigma_\rho(x,z)\bar D_F(z,x) \bigr] \,. \nonumber 
\end{align}
Note that we have expressed the integration limits in terms of the usual
$\Theta$-function.
It is helpful to absorb $\Theta(x^0-z^0)$ into retarded and advanced
propagators,
\begin{eqnarray}
  D_R(x,y) & \equiv & \Theta(x^0-y^0)D_\rho(x,y) \,, \\
  D_A(x,y) & \equiv & -\Theta(y^0-x^0)D_\rho(x,y) \,,
\end{eqnarray}
again with analogous definitions for $\bar D$ and the self-energies.
The Wigner representation of the various two-point functions reads 
\begin{equation}\label{WignerTrafoGeneral}
  A(X,p)  =  (-i)^\rho{\textstyle \int} d^4 s \, e^{ips}\, A(X+s/2,X-s/2)\,,
\end{equation}
where $\rho=1$  for $A=D_\rho,\bar D_\rho,\Sigma_\rho,\bar\Sigma_\rho$, and zero
otherwise.
Using Eq.\,\eqref{GradExp} and assuming $t_{init}\to-\infty$,
the Wigner representation of Eq.\,\eqref{DivJ_KB} is
\begin{align}\label{DivJ_KB_Wigner}
    {\cal D}^\mu j_\mu(x)  = & -i {\textstyle \int } {d\Pi_p}   \,  \\
         & \times \! \bigl[  e^{-i\diamondsuit} \{\Sigma_F\}\{D_A\} +
e^{-i\diamondsuit} \{\Sigma_R\}\{D_F\} \nonumber\\
                 & -  e^{-i\diamondsuit} \{\bar\Sigma_F\}\{\bar D_A\} -
e^{-i\diamondsuit} \{\bar \Sigma_R\}\{\bar D_F\} \bigr] \,. \nonumber
\end{align}
All two-point functions are evaluated at the point $(X,p)$ in phase-space.
The gradient expansion formally follows by expanding the exponentials
in powers of the derivative operator $\diamondsuit \propto \partial_X$ defined
in Eq.\,\eqref{PB}.

\subsection{Zeroth order}

Evaluating Eq.\,\eqref{DivJ_KB_Wigner} at zeroth order in the gradient expansion
yields
\begin{align}
   {\cal D}^\mu j_\mu(x)|_0   = -i {\textstyle \int } {d\Pi_p}  \, \bigl\{ &
\Sigma_F^0 D_A + \Sigma_R^0 D_F  \nonumber\\
                               {} - & \bar\Sigma_F^0\bar D_A - \bar
\Sigma_R^0\bar D_F \bigr\} \,.
\end{align}
By substituting $p\rightarrow -p$ in the second line and using
\begin{eqnarray}\label{DDbar}
   D_{(F,A,R,\gtrless)}(X,p) & = & \bar D_{(F,R,A,\lessgtr)}(X,-p) \,, \\
   D_{\rho}(X,p)    & = & -\bar D_{\rho}(X,-p) \nonumber\\
                    & = & -i[D_R(X,p)-D_A(X,p)]\,, \nonumber
\end{eqnarray}
this equation can be simplified to
\begin{equation}
   {\cal D}^\mu j_\mu(x)|_0   =  - {\textstyle \int {d\Pi_p} } 
 \, \bigl[ \Sigma_F^0 D_\rho - \Sigma_\rho^0 D_F \bigr] \,.
\end{equation}
Using Eq.\,\eqref{FRho} to re-express this result in terms of Wightman functions
(note factors $i$ in Eq.\,\eqref{WignerTrafoGeneral}) yields
\begin{align}
   {\cal D}^\mu j_\mu(x)|_0   =  - {\textstyle \int {d\Pi_p} }  \, 
\bigl[ \Sigma_<^0 D_> - \Sigma_>^0 D_< \bigr] \,.
\end{align}
Finally, inserting the identity $1 = \Theta(p_0) + \Theta(-p_0)$ and using
Eq.\,\eqref{DDbar} yields,
\begin{align}
   {\cal D}^\mu j_\mu(x)|_0   =  - {\textstyle \int {d\Pi_p} \Theta(p_0)}  \,
\bigl[ & \Sigma_<^0 D_> - \Sigma_>^0 D_< \nonumber\\
    - & \bar\Sigma_<^0 \bar D_> + \bar \Sigma_>^0 \bar D_<\bigr] \,,
\end{align}
which corresponds to Eq.\,\eqref{DivJ_Boltzmann}.

\subsection{First order}

The first-order gradient contribution to Eq.\,\eqref{DivJ_KB_Wigner}
consists of two parts. One obviously involves the linear term in the 
expansion of $e^{-i\diamondsuit}$.
In addition, it is important to realize that the self-energies, see
Eq.\,\eqref{Sigmac}, contain an
internal toy-neutrino line described by the out-of-equilibrium propagator
$G^{ij}$.
In order to obtain a consistent gradient expansion, it is important to perform
also a gradient
expansion of the equation of motion for $G^{ij}$,
which is discussed in Appendix~\ref{app:G1}. This means we also have to expand
the self-energies,
\begin{equation}
  \Sigma = \Sigma^0 + \Sigma^1 + \dots \,.
\end{equation}

Doing similar manipulations as for the zeroth-order equation yields
\begin{eqnarray}\label{DivJ_1}
  {\cal D}^{\mu} j_\mu|_1 & = &  -{\textstyle \int{d\Pi_p} } \big[ 
\Sigma_<^1D_> - \Sigma_>^1D_< \\
                          &   & {} +2 \diamondsuit\{\Sigma_F^0\}\{D_h\} +2
\diamondsuit\{\Sigma_h^0\}\{D_F\} \big] \,, \nonumber
\end{eqnarray}
where $D_h(X,p) \equiv [D_R(X,p)+D_A(X,p)]/2$.
Inserting again $1 = \Theta(p_0) + \Theta(-p_0)$
and using Eq.\,\eqref{DDbar} yields Eq.\,\eqref{DivJ_FirstOrder}.

\section{\label{app:G1}Gradient expansion for real scalar fields}
The dynamics of the system 
of real scalar fields is described by the non-equilibrium 
generalization of the Schwinger--Dyson equation \cite{Garny:2009qn}
\begin{align}
\label{SchwingerDyson}
(G^{-1})^{ij}(x,y)=(\mathscr{G}^{-1})^{ij}(x,y)-\Pi^{ij}(x,y)\,,
\end{align}
where $G^{ij}$ is the full dressed  propagator  of the ``heavy neutrinos'', 
$\mathscr{G}^{ij}$ is the diagonal propagator of the free fields and 
$\Pi^{ij}$ is the self-energy. Let us now split the self-energy matrix  
$\hat \Pi$ into the diagonal, $\hat \varPi$, and off-diagonal, 
$\hat \varPi^{'}$, components and introduce a diagonal propagator 
$\hat {\cal G}$ defined by the equation 
\begin{align}
\label{DiagonalEq}
\hat {\cal G}^{-1}(x,y)=\hat{\mathscr{G}}^{-1}(x,y)-\hat \varPi(x,y)\,.
\end{align}
Subtracting Eq.\,\eqref{DiagonalEq} from Eq.\,\eqref{SchwingerDyson} we find:
\begin{align}
\label{OffDiagDiagonalEq}
\hat G^{-1}(x,y) = \hat {\cal G}^{-1}(x,y)-\hat \varPi^{'}(x,y)\,.
\end{align}
Multiplying Eq.\,\eqref{OffDiagDiagonalEq} by $\hat G$ from the left, by 
$\hat {\cal G}$ from the right and integrating over the closed-time-path 
contour \cite{Schwinger:1960qe,Keldysh:1964ud} we obtain a formal solution 
for the full non-equilibrium propagator \cite{Garny:2009qn}:
\begin{align}
\label{GgtrlessFormalSol}
\hat G_\gtrless(x,y)= \hat {\cal G}_\gtrless(x,&\,y)-{\textstyle\iint}
\mathscr{D}^4u \mathscr{D}^4v\, \theta(u^0) \, \theta(v^0)\nonumber\\
\times \bigl[ & \hat G_R(x,u) \hat \varPi^{'}_\gtrless(u,v) \hat {\cal
G}_A(v,y)\nonumber\\
+ & \hat G_\gtrless(x,u)\hat \varPi^{'}_A(u,v)\hat {\cal G}_A(v,y)\nonumber\\
+ & \hat G_R(x,u)\hat \varPi^{'}_R(u,v)\hat {\cal G}_\gtrless(v,y)\bigr]\,.
\end{align}
If the mass spectrum of the heavy scalars is strongly hierarchical,
i.e.~$M_1^2 \ll M_2^2$, one can approximate the full propagators $\hat G$ on 
the right-hand side of Eq.\,\eqref{GgtrlessFormalSol} by the corresponding 
diagonal propagators $\hat{\cal G}$. That is, in this approximation the 
dynamics of the diagonal and off-diagonal components of $\hat G$ is 
completely determined by the dynamics of $\hat {\cal G}$. Wigner-transforming 
the resulting  expression we obtain to leading order in the gradients
$\hat G_\gtrless=\hat G_\gtrless^{0}+\hat G_\gtrless^{1}$,
where
\begin{subequations}
\label{Ggtrless}
\begin{align}
\label{Ggtrless0}
\hat G_\gtrless^{0} \simeq \,\, \hat {\cal G}_\gtrless & -
\bigl[\hat {\cal G}_R \hat \varPi^{'}_\gtrless \hat {\cal G}_A
+\hat {\cal G}_\gtrless\hat \varPi^{'}_A\hat {\cal G}_A+\hat {\cal G}_R\hat
\varPi^{'}_R\hat {\cal G}_\gtrless\bigr]\\
\label{Ggtrless1}
\hat G_\gtrless^{1} \simeq -i
\bigl[ & \diamondsuit\{ \hat {\cal G}_R,\hat \varPi^{'}_\gtrless, \hat {\cal
G}_A \}\nonumber\\
+ & \diamondsuit \{ \hat {\cal G}_\gtrless, \hat \varPi^{'}_A,\hat {\cal G}_A\}+
 \diamondsuit \{ \hat {\cal G}_R,\hat \varPi^{'}_R,\hat {\cal
G}_\gtrless\}\bigr]\,.
\end{align}
\end{subequations}
The generalized derivative operators in Eq.\,\eqref{Ggtrless1} are defined by
\begin{align}
\diamondsuit\{A,B,C\} \equiv \diamondsuit\{A\}\{BC\} +
A\diamondsuit\{B\}\{C\}\,.
\end{align}

A system in thermal equilibrium is stationary. Therefore, in equilibrium 
the right-hand side of Eq.\,\eqref{Ggtrless1} vanishes. One can expect that 
in the early universe this contribution is proportional to the expansion
rate $\hubble$ of the universe. A substitution of Eq.\,\eqref{Ggtrless1} into
the 
expressions for the Wigner-transforms of the self-energy, Eq.\,\eqref{SigmasWT},
gives us the first-order corrections  $\Sigma^1_\gtrless$ to the self-energies.
 
\section{Derivation of the additional source term}\label{app:Delta1}

In this appendix we derive the additional \CP-violating source term ${\cal
S}_{\dot T}(\invt)$
. Our starting point is the CTP evolution equation for the $B-L$ current 
\eqref{DivJ_KB_Derivation}. The additional source term arises at
first order in the gradient expansion, see Eq.\,\eqref{DivJ_1}.
Inserting baryon-symmetric
propagators, $D_\gtrless,\bar D_\gtrless \rightarrow D^{sym}_\gtrless$
as well as the self-energy given in Eq.\,\eqref{SigmasWT}
into Eq.\,\eqref{DivJ_1} yields
\begin{align}
\label{S1}
\lefteqn{ {\cal S}(\invt)|_1 \equiv K\cdot{\cal D}^\mu j_\mu(x)|_{1,sym}  \simeq
}\nonumber\\
& 2K\,\Im(g_i^*g_j) {\textstyle \int d\Pi_p \int d\Pi_q \int d\Pi_k }
\Theta(p_0)  (2\pi)^4  \delta(k-p-q) \nonumber\\ 
& \times \bigl[ \Im(G_>^{1,ij})D^{sym}_<D^{sym}_< 
 - \Im(G_<^{1,ij})D^{sym}_>D^{sym}_> \bigr]. 
\end{align}
Here we have neglected the contributions in the second line of
Eq.\,\eqref{DivJ_1},
similar as in~\cite{Prokopec:2004ic}, and the first-order toy-neutrino
propagators $G^1$ are given by Eq.\,(\ref{Ggtrless1}). For brevity, we suppress
the superscript of $D^{sym}$ in the following.

Next, we set all particle species into local thermal equilibrium
in order to obtain ${\cal S}_{\dot T}={\cal S}|_1^{LTE}$.
This means that the propagators fulfill the Kubo-Martin-Schwinger (KMS) relation
\begin{eqnarray}
	D_>^{LTE}(X,k) & = & e^{\beta (k\cdot u)}D_<^{LTE}(X,k) \,,
\end{eqnarray}
where $u=(1,0,0,0)$ in the co-moving frame of the thermal bath, and
$\beta=\beta(t) \equiv 1/T(t)$. Note that this condition
implies that $f_b=1/[e^{\beta (k\cdot u)}-1]$, see Eq.\,\eqref{Dgtrless}.
For the toy-neutrino, the condition of \emph{local} equilibrium implies
that the \emph{zero}-order propagators $G_\gtrless^{0}$ fulfill KMS.
In particular, this implies that
\begin{eqnarray}
	{\cal G}_>^{LTE}(X,k) & = & e^{\beta (k\cdot u)}{\cal G}_<^{LTE}(X,k) \,.
\end{eqnarray}
Again, in the following we suppress the superscript.
These relations greatly simplify Eq.\,\eqref{S1} after inserting
Eq.\,(\ref{Ggtrless1}).
As an example, we consider the following contribution:
\begin{align}
	\diamondsuit & \{ {\cal G}_>^{ii}, \Pi^{ij}_A,{\cal G}_A^{jj}\} D_< D_< - 
     \diamondsuit\{ {\cal G}_<^{ii}, \Pi^{ij}_A,{\cal G}_A^{jj}\} D_> D_> =
\nonumber\\
	 = & [ (\partial_{k^\alpha} {\cal G}_>^{ii}) D_< D_< - (\partial_{k^\alpha}
{\cal G}_<^{ii}) D_> D_> ] \times {\cal D}^\alpha \Pi^{ij}_A{\cal G}_A^{jj}
\nonumber\\
	   {} - &  [ {\cal D}^\alpha ( {\cal G}_>^{ii}) D_< D_< - ({\cal D}^\alpha
{\cal G}_<^{ii}) D_> D_> ] \times \partial_{k^\alpha} \Pi^{ij}_A{\cal G}_A^{jj}
\nonumber\\
	 {} + & [  {\cal G}_>^{ii} D_< D_< -  {\cal G}_<^{ii} D_> D_> ] \times
\diamondsuit\{ \Pi^{ij}_A \}\{ {\cal G}_A^{jj} \} \,.
\end{align}
Using the LTE relations, we see that the third line cancels. Furthermore, using
again LTE, we find
\begin{align}
	\partial_{k^\alpha} {\cal G}_>^{ii} & = \partial_{k^\alpha} [ e^{\beta (k\cdot
u)} {\cal G}_<^{ii} ] = \beta u_\alpha e^{\beta (k\cdot u)} {\cal G}_<^{ii}
\nonumber\\
 & \hspace{4cm}+ e^{\beta (k\cdot u)} \partial_{k^\alpha} {\cal G}_<^{ii} \,,\\
	{\cal D}^\alpha {\cal G}_>^{ii}  &  = {\cal D}^\alpha [ e^{\beta (k\cdot u)} {\cal
G}_<^{ii} ] = ({\cal D}^\alpha\beta) k\cdot u\, e^{\beta (k\cdot u)} {\cal
G}_<^{ii} \nonumber\\
 & \hspace{4cm} + e^{\beta (k\cdot u)} {\cal D}^\alpha {\cal G}_<^{ii} \,.
\end{align}
Inserting these relations in the first and second line of the above equation, we
find that there are again some cancellations,
\begin{align}
	& \diamondsuit \{ {\cal G}_>^{ii}, \Pi^{ij}_A,{\cal G}_A^{jj}\} D_< D_< 
    - \diamondsuit \{ {\cal G}_<^{ii}, \Pi^{ij}_A,{\cal G}_A^{jj}\} D_> D_>
=\nonumber\\
	& =  {\cal G}_<^{ii} D_> D_> \, [(\beta u\cdot{\cal D} - ({\cal
D}^\alpha\beta) k\cdot u\, \partial_{k^\alpha})(\Pi^{ij}_A{\cal G}_A^{jj})] \,.
\end{align}
Similar simplifications are obtained for the other two first-order gradient
terms of $G_\gtrless^{1,ij}$,see Eq.\,(\ref{Ggtrless1}).
To shorten the notation we introduce the differential operator
\begin{equation}
	\tilde {\cal D}^\alpha \equiv {\cal D}^\alpha - \frac{1}{\beta} k^\alpha
({\cal D}^\sigma\beta) \partial_{k^\sigma} \,. 
\end{equation}
Then, putting everything together, yields
\begin{align}
  &{\cal S}(\invt)|_{\dot T} = {\cal S}(\invt)|_1^{LTE}  \nonumber\\
 & = -  K\,\Im(g_i^*g_j)
{\textstyle \int} d\Pi_p  d\Pi_k d\Pi_q \, (2\pi)^4  \delta(k-p-q)\Theta(p_0)
\nonumber\\ 
                                &    \times \beta u_\alpha \bigl[ \Im [\tilde
{\cal D}^\alpha(i\Pi^{ij}_A{\cal G}_A^{jj})] \, {\cal G}_>^{ii}D_<D_<
\nonumber\\
                                &    - \Im[\tilde {\cal D}^\alpha(i{\cal
G}_R^{ii}\Pi^{ij}_R)] \, {\cal G}_>^{jj}D_<D_< \nonumber\\
                                &   + \Im(i{\cal G}_R^{ii}\tilde{\cal
D}^\alpha{\cal G}^{jj}_A - i \tilde{\cal D}^\alpha{\cal G}_R^{ii}{\cal
G}^{jj}_A) \,\Pi_>^{ij}D_<D_< \bigr] \,.
\end{align}
The parts where the space-time derivative acts on a retarded or advanced
propagator ${\cal G}_{R(A)}$
are suppressed, since they only depend on temperature via the thermal mass,
which is of higher order
in the coupling within the toy-model. Therefore, we neglect the third line. In
addition, using relations
between retarded and advanced quantities and interchanging $i \leftrightarrow j$
in the second line gives
\begin{align}
	{\cal S}(\invt)|_{\dot T}   =  2K\, & |g_i|^2{\textstyle \int}
d\Pi_pd\Pi_kd\Pi_q \, (2\pi)^4 \delta(k-p-q)\Theta(p_0)  \nonumber\\
	                                   \times & \epsilon^i_{\dot T}(k,T) \left[{\cal
G}^{ii}_<D_>D_> + {\cal G}^{ii}_>D_<D_<\right]\,,
\end{align}
where
\begin{eqnarray}
	\epsilon^i_{\dot T}(k,T) = {\textstyle \frac{1}{2}} \Im(g_j/g_i) \beta u_\alpha \tilde
{\cal D}^\alpha \Im[i{\cal G}_R^{jj}(k)\Pi_R^{ji}(k)].
\end{eqnarray}
In the hierarchical limit, one obtains for the lightest toy-neutrino ($i=1$):
\begin{eqnarray}
	\epsilon_{\dot T}(k,T) \equiv \epsilon^{i=1}_{\dot T}(k,T) = - \epsilon^{vac} \, \frac{\beta}{2}
u_\alpha \tilde {\cal D}^\alpha L_h(X,k),
\end{eqnarray}
where $\epsilon^{vac}$ is the \CP-violating parameter in vacuum.
The loop integral $L_h$ parameterizes the real part of the self-energy of the
heavy real fields, see \cite{Garny:2009qn},
\begin{align}
	 L_h(X,k) & = 16\pi {\textstyle\int} d\Pi_q \, D_F(q;T) D_h(k-q) \nonumber\\
                   & \equiv L_h(k;T(t)) \,.
\end{align}
In the co-moving frame:
\begin{align}
	u_\alpha \tilde {\cal D}^\alpha & = \tilde {\cal D}^0 = {\cal D}^0 -
\frac{\dot\beta}{\beta} k^0 \partial_{k^0} = \dot T \frac{\partial}{\partial T}
+ \frac{\dot T}{T}k^0 \partial_{k^0} \nonumber\\
&  = -{\hubble} \left[ T \frac{\partial}{\partial T} + k^0 \partial_{k^0}
\right]\,.
\end{align}
This result can easily be generalized to a frame that is boosted with respect to
the co-moving frame:
\begin{equation}
	u_\alpha \tilde {\cal D}^\alpha = -{\hubble} \left[ T \frac{\partial}{\partial
T} + (u\cdot k) (u \cdot \partial_{k}) \right]\,.
\end{equation}
Thus:
\begin{align}
\hspace{-2mm}
	\epsilon_{\dot T}(k,T) = \epsilon^{vac} & \times \frac{\beta {\hubble}}{2}  \left[
T\frac{\partial}{\partial T} +  (u\cdot k) (u \cdot \partial_{k})\right]
\nonumber\\ 
        & \times L_h(k;T).
\end{align}
In the quasi-particle limit, the loop integral is given by
\begin{align}
	L_h(k;T) = 16 & \pi {\textstyle \int} d\Pi_q^3 \, \Big( \left[ {\textstyle
\frac12} + f_b(q\cdot u;T) \right] \frac{-{\cal P}}{(k-q)^2-m^2} \nonumber\\
	            + & \left[ {\textstyle \frac12} + f_{\bar b}(q\cdot u;T) \right]
\frac{-{\cal P}}{(k+q)^2-m^2} \Big) \,,
\end{align}
where $q^2=m^2, q^0=\sqrt{m^2+\vec{q}^2}$ and $f_b\simeq f_{\bar b}=f^{BE}$
for an approximately symmetric medium. The temperature-derivative can be
evaluated using the result in \cite{Garny:2009qn}:
\begin{align}
T\frac{\partial}{\partial T} L_h(k;T) =  \frac{1}{\pi|\vec{k}|} \int_0^\infty &
dE \; \left[T\frac{\partial}{\partial T} f_b^{BE}(E;T) \right] \nonumber \\
\times & \ln \left| \frac{ (2E+|\vec{k}|)^2 - k^2 }{ (2E-|\vec{k}|)^2 - k^2 }
\right| \,,\nonumber
\end{align}
where $k^2=M^2$ and $m\approx 0$. 
The derivative with respect to the momentum is given by
\begin{align}
	\frac{\partial}{\partial k}& L_h(k;T)  =  16\pi {\textstyle\int} d\Pi_q^3 \,
\left[ {\textstyle \frac12} + f_b^{BE}(q\cdot u;T) \right]
      \nonumber\\
   & \times  {\cal P} \left[ \frac{2(k-q)}{[(k-q)^2-m^2]^2} +
\frac{2(k+q)}{[(k+q)^2-m^2]^2} \right] \,.
\end{align}
This expression is covariant, i.e.~valid in any frame. In the co-moving frame,
where $u=(1,0,0,0)$, we need to compute only the derivative with respect to
$k^0$. In a general frame, this corresponds to $u \cdot \partial_k$. Thus,
\begin{align}
	k^0&\partial_{k^0}L_h(k;T)|_{comoving}  =   (u\cdot k) \times u\cdot
\frac{\partial}{\partial k} L_h(k;T) \nonumber\\
	                                       = & 16\pi \, (u\cdot k) {\textstyle
\int} d\Pi_q^3 \, \left[{\textstyle \frac12} + f_b^{BE}(q\cdot u;T) \right]
\nonumber\\
& \times  {\cal P} \left[ \frac{2u\cdot (k-q)}{[(k-q)^2-m^2]^2}  + \frac{2u\cdot
(k+q)}{[(k+q)^2-m^2]^2} \right] \,.
\end{align}
The above integral contains a ``vacuum'' and a ``medium'' part, where the latter
is proportional to $f_b^{BE}(q\cdot u;T)$.
An explicit calculation shows that both integrals are well-defined (no UV, IR or
on-shell-pole divergences), and yields (for $m=0$)
\begin{align}
\hspace{-2mm}
	(u\cdot & k)  (u\cdot \partial_k) \, L_h^{vac}(k;T)  = 
\frac{E_k}{2\pi|\vec{k}|} \ln \left| \frac{E_k-|\vec{k}|}{E_k+|\vec{k}|} \right|
\,, \\
	(u\cdot & k)   (u\cdot \partial_k) \, L_h^{med}(k;T)  = -
\frac{E_k}{2\pi|\vec{k}|}\nonumber\\
& \times  \int_0^\infty \!\!\! dE \frac{\partial f_b^{BE}}{\partial E} \ln
\left| \frac{M^4 - 4E^2(E_k-|\vec{k}|)^2}{M^4 -4E^2(E_k+|\vec{k}|)^2}
\right|\,. 
\end{align}
Here $|\vec{k}|$ is the momentum of the decaying particle in the comoving frame
(rest-frame of the medium), and
$E_k=\sqrt{M^2+\vec{k}^2}$ is its energy.
For the part containing the temperature-derivative of $L_h$, we use
\begin{equation}
	T\frac{\partial}{\partial T} f_b^{BE}(E;T) = - E\frac{\partial}{\partial E}
f_b^{BE}(E;T) \,.
\end{equation}
Then one obtains
\begin{align}
    \bigg[ T\frac{\partial}{\partial T}   +  (u\cdot k) (u \cdot &
\partial_{k})\bigg] L_h(k;T) =  
   \frac{E_k}{2\pi|\vec{k}|} \Bigg[ \ln \left|
\frac{E_k-|\vec{k}|}{E_k+|\vec{k}|} \right|\nonumber \\
	     - \int_0^\infty \!\!\! dE \frac{\partial f_b^{BE}}{\partial E} \Bigg\{ &
\ln \left| \frac{M^4 - 4E^2(E_k-|\vec{k}|)^2}{M^4 -4E^2(E_k+|\vec{k}|)^2}
\right| \nonumber\\
	     +  \frac{2E}{E_k} & \ln \left| \frac{ (2E+|\vec{k}|)^2 - M^2 }{
(2E-|\vec{k}|)^2 - M^2 } \right| \Bigg\} \Bigg] \,.
\end{align}
It is easy to convince oneself that the remaining integral over the energy $E$
is free of UV or IR divergences, and that the
integral exists in the vicinity of the zeros of the arguments of the logarithm
occurring inside the integration region. 
Thus, we finally arrive at the result
\begin{align}
\label{deltaFull}
	\epsilon_{\dot T}(k,T) & =  \epsilon^{vac} \times  \frac{\beta {\hubble}}{2}  \times 
\frac{E_k}{2\pi|\vec{k}|} \Bigg( \ln \left| \frac{E_k-|\vec{k}|}{E_k+|\vec{k}|}
\right| \nonumber\\
	     &- \int_0^\infty \!\!\! dE \frac{\partial f_b^{BE}}{\partial E} \Bigg\{ 
\ln \left| \frac{M^4 - 4E^2(E_k-|\vec{k}|)^2}{M^4 -4E^2(E_k+|\vec{k}|)^2}
\right|
\nonumber\\
	           & + \frac{2E}{E_k} \ln \left| \frac{ (2E+|\vec{k}|)^2 - M^2 }{
(2E-|\vec{k}|)^2 - M^2 } \right| \Bigg\} \Bigg) \,.
\end{align}
In the non-relativistic limit $|\vec{k}| \sim T \ll M$, the 
main contribution to the integration over $E$ comes from the region $E \ll M$,
due to the exponential suppression in the Bose-Einstein function. Therefore, we
may expand the logarithms in the above expression for $|\vec{k}|, E \ll M$:
\begin{eqnarray}
	\ln \left| \frac{E_k-|\vec{k}|}{E_k+|\vec{k}|} \right| & \rightarrow &
-2|\vec{k}|/E_k \,, \\
	\{ \dots \}                                            & \rightarrow & 16
E^2|\vec{k}|^3/(M^4E_k) \,.
\end{eqnarray}
Then it is easy to perform the energy integral:
\begin{equation}
	\int_0^\infty \!\!\! dE \frac{\partial f_b^{BE}}{\partial E} E^2 = -
\int_0^\infty \!\!\! dE f_b^{BE} 2E = -\frac{T^2\pi^2}{3} \,.
\end{equation}
Using this, we obtain the leading contributions in the non-relativistic limit,
\begin{equation}
	\epsilon_{\dot T}(k,T) \rightarrow \epsilon^{vac} \times \frac{\beta {\hubble}}{2} 
\left( -\frac{1}{\pi} + \frac{8\pi T^2\vec{k}^2}{3M^4} \right) \,.
\end{equation}
The second contribution in the brackets is suppressed for $T<M$, so that we finally obtain
\begin{equation}
	\epsilon_{\dot T}(k,T) \simeq  - \epsilon^{vac} \times\frac{{\hubble}}{2\pi T} \,.
\end{equation}
\vfill

\end{appendix}



\end{document}